\documentclass[twocolumn,preprintnumbers,amsmath,amssymb,superscriptaddress]{revtex4}
\usepackage{epsfig,bm}
\usepackage{ulem}
\usepackage{amsmath,color}

\begin{document}
\title{Nuclear surface diffuseness revealed in
nucleon-nucleus diffraction}
\author{S. Hatakeyama}
\affiliation{Department of Physics,
  Hokkaido University, Sapporo 060-0810, Japan}
\author{W. Horiuchi}
\affiliation{Department of Physics,
  Hokkaido University, Sapporo 060-0810, Japan}
\author{A. Kohama}
\affiliation{RIKEN Nishina Center, RIKEN, Wako-shi, Saitama 351-0198, Japan}

\begin{abstract}
  Nuclear surface provides useful information on
  nuclear radius, nuclear structure as well as properties
  of nuclear matter.
  We discuss the relationship between
  the nuclear surface diffuseness
  and elastic scattering differential cross section
  at the first diffraction peak of high-energy nucleon-nucleus scattering
  as an efficient tool
  in order to extract the nuclear surface information
  from limited experimental data involving short-lived unstable nuclei.
  The high-energy reaction is described 
  by a reliable microscopic reaction theory, the Glauber model.  
  Extending the idea of the black sphere model,
  we find one-to-one correspondence between
  the nuclear bulk structure information
  and proton elastic scattering diffraction peak.
  This implies that we can extract
  both the nuclear radius and diffuseness simultaneously,
  using the position of the first diffraction peak and 
  its magnitude of the elastic scattering differential 
  cross section.
  We confirm the reliability of this approach
  by using realistic density distributions
  obtained by a mean-field model.
\end{abstract}
\maketitle

\section{Introduction}

A nucleus is composed of proton and neutron interacting 
via nuclear force. They make a self-consistent-mean field
that results in forming the nuclear shell structure.
A systematic analysis of nuclear charge radii
via the electron-elastic scattering
have revealed that nuclei have saturated internal density
and relatively sharp surface that defines a nuclear 
radius~\cite{deVries87}.
Advances in the radioactive ion beam facilities
have made us possible to study properties
of short-lived unstable nuclei.
From such facilities, 
exotic structure of neutron-rich unstable nuclei
was reported,
which has never been observed in stable nuclei,
e.g., halo~\cite{Tanihata85} and 
developed skin~\cite{Suzuki95} structure.

Such exotic structure is dominated by
nuclear dynamics at around the nuclear surface.
For example, a nuclear deformation plays a crucial
role in enlarging the nuclear radius,
because it drastically changes the density profile at around
the nuclear surface,
and has actually been confirmed by 
the systematic analyses of
the total reaction cross sections on a carbon target
~\cite{Takechi10, Minomo11,Minomo12,Sumi12,Horiuchi12,
Watanabe14,Takechi14}.
Also, it is found that excitations of
the outermost single-particle-neutron orbits
play an essential role to determine
the low-lying electric-dipole strengths of neutron-rich
isotopes~\cite{Inakura13,Inakura14,Ebata14,Horiuchi17a}.

Since the density profile at around the nuclear surface is
a rich source of the nuclear structure information,
a systematic investigation of the nuclear surface density distributions
must be worth studying.
However, the neutron density distribution
is difficult to probe by the traditional electron scattering.
Alternatively, the proton-elastic scattering is suitable for that purpose
~\cite{Sakaguchi17}.
Recent precise measurements up to large scattering angles
were successful in extracting the neutron density distributions
of Sn and Pb isotopes with the help of known
proton density distributions~\cite{Terashima08,Zenihiro10}.
To apply it for unstable nuclei,
such measurement in the inverse kinematics is useful but it is
not easy to obtain the cross sections at large scattering angles
because most of incident particles are scattered in
the forward angles at high incident energies.
Since precise experimental cross sections
are limited to small scattering angles, 
it is convenient to know what information we
can obtain from limited cross section data.

In this paper, we 
perform a ``numerical experiment'' systematically
  using theoretically obtained nucleon-nucleus scattering cross sections
  focusing on the reactions of small scattering angles up to a few
  diffraction peaks to see
which extent the information on the nuclear surface can be obtained.
For this purpose, we start with an idea
of a simple black sphere (BS) picture,
which assumes a nucleus is a completely absorptive object
at a sharp-cut-square-well radius.
The model is mathematically equivalent to 
the Fraunhofer diffraction model~\cite{Bethe,Glauber}, 
which offers one-to-one correspondence between
the nuclear radius and
the diffraction peak position.
Though it is not perfect,
the idea can be a zeroth order approximation
of the proton-nucleus scattering remarking the fact that
the BS model explains fairly well a systematic trend of the
proton-nucleus total reaction cross sections,
which was originally pointed out
by Kohama {\it et al.}~\cite{Kohama04,Kohama05,Kohama16}.

In reality, since the total nucleon-nucleon cross section
is not large enough at medium- and high-incident energies,
the proton-nucleus scattering is not completely
absorptive at around the nuclear surface where
the nuclear density is not well saturated.
A simple model approach 
based on the BS picture and the proton optical depth
shows that the effect of the surface diffuseness
plays an essential role to determine
the incident energy dependence
of the total reaction cross sections
of the proton-nucleus scattering~\cite{BS3,Kohama16}.
Here we discuss the role of the nuclear transparency due to
the surface diffuseness which is not explicitly
taken into account in the BS model 
by comparing it with a microscopic high-energy
reaction theory, the Glauber model~\cite{Glauber}.
To understand the significance of its good reproducibility 
of the data using such a phenomenology is 
another purpose of this paper.

This paper is organized as follows:
In the next section, we briefly explain
calculations of the elastic scattering differential cross section
of the high-energy nucleon-nucleus scattering in the Glauber model.
In Sec.~\ref{Fermi.sec}, we demonstrate how
the elastic scattering differential cross section reflects
the density profile at the nuclear surface.
A systematic analysis is performed by using a two-parameter 
Fermi (2pF) distribution as the density profile
which clearly defines the nuclear ``diffuseness''.
We find one-to-one correspondence
between the nuclear diffuseness and 
the magnitude of the cross section at the first
peak position.
To quantify the sensitivity to the density profile
at around the surface region,
we investigate, in Sec.~\ref{samp.sec}, 
the spatial distributions of the scattering amplitude
at the first and second peak positions
of the elastic scattering differential cross sections.
In Sec.~\ref{extdiff.sec},
we propose a simple way to extract both the nuclear radius
and diffuseness information from
the elastic scattering differential cross sections
for future application to short-lived unstable nuclei.
We extract the information on
the surface diffuseness of density distributions
obtained via a microscopic mean field approach.
By assuming the 2pF density distributions,
the unknown ``diffuseness'' and nuclear radius
are uniquely determined in such a way that 
the first peak position and its magnitude of
the elastic scattering differential cross section
are reproduced simultaneously.
It will be convenient to know the first peak position before measurement
from other observables in which we show, in Sec.~\ref{BS.sec},
a relationship between the first peak position and the total reaction
cross section with the help of the BS model.
The possibility of extracting proton and neutron diffuseness 
separately is discussed in Sec.~\ref{diffpn.sec}.
Conclusions are presented in Sec.\ref{conclusion.sec}.

\section{Elastic scattering differential cross section 
in the Glauber model}
\label{Glauber.sec}

The Glauber model is a microscopic theory which
is widely used to study high-energy nucleus-nucleus 
collisions~\cite{Glauber}.
With the help of the adiabatic and eikonal approximations,
the final state scattering wave function
is greatly simplified as
\begin{align}
  \left|\Phi_f\right>=e^{i\chi}\left|\Phi_i\right>,
\end{align}
where $e^{i\chi}$ is the so-called phase-shift function,
which includes all information of the high-energy nuclear collision.
The elastic scattering differential cross section can be calculated by
\begin{align}
\frac{d\sigma}{d\Omega}(\theta)=|F(\theta)|^2
\end{align}
with the elastic scattering amplitude 
\begin{align}
  F(\theta)=\frac{iK}{2\pi}\int e^{-i\bm{q}\cdot\bm{b}}
  \left(1-e^{i\chi(\bm{b})}\right)\,d\bm{b},
\end{align}
where $K$ is the wave number in the relativistic kinematics,
$\bm{q}$ the momentum transfer vector,
and $\bm{b}$ is the impact parameter
vector perpendicular to the beam direction ($z$),
and thus $\bm{q}\cdot\bm{b}=2Kb\sin\frac{\theta}{2}$.
Evaluation of $e^{i\chi(\bm{b})}$ is in general difficult
because it involves multiple integration~\cite{Glauber}.
Though it could be possible to perform the integration
by using a Monte Carlo technique~\cite{Varga02,Nagahisa18}
or a factorization procedure by using a Slater determinant
wave function~\cite{Bassel68, Ibrahim10,Hatakeyama14,Hatakeyama15},
we, however, employ the optical-limit approximation (OLA)
for the sake of simplicity.
The phase-shift function of the OLA is given
as the leading order of the cumulant expansion of 
the full phase-shift function~\cite{Glauber,Suzuki03}
\begin{align}
  i\chi(\bm{b})
  \simeq -\sum_{N=p,n}\int \rho_N(\bm{r})\Gamma_{pN}(\bm{b}-\bm{s})\,d\bm{r},
\end{align}
where $\bm{r}=(\bm{s},z)$ with $\bm{s}$
being a two-dimensional vector perpendicular to $z$.
Inputs to the theory are density distributions $\rho_N(\bm{r})$
of proton ($N=p$) and neutron ($N=n$),
and the proton-nucleon profile function $\Gamma_{pN}$.
  As exemplified
  in Refs.~\cite{Varga02, Ibrahim10, Hatakeyama14,Hatakeyama15,Nagahisa18},
  the OLA works well for many cases of nucleon-nucleus scattering.
  The multiple scattering effect would be neglected
  and even becomes smaller for systems
  involving medium to heavy nuclei
 as was shown in Refs.~\cite{Hatakeyama14,Hatakeyama15}.

The nucleon-nucleon profile function at incident energy per nucleon $E$
is usually parametrized as~\cite{Ray79}
\begin{align}
  \Gamma_{NN}(\bm{b},E)=\frac{1-i\alpha_{NN}(E)}{4\pi\beta_{NN}(E)}
  \sigma_{NN}^{\rm tot}(E)
  \exp\left[-\frac{\bm{b}^2}{2\beta_{NN}(E)}\right],
\label{profile.eq}
\end{align}
where $\alpha_{NN}$ is the ratio of the real and imaginary parts
of the scattering amplitude at the forward angle,
and $\beta_{NN}$ is a slope parameter.
For the sake of simplicity, we first use averaged $NN$ profile function
given in Ref.~\cite{Horiuchi07} for most of discussions 
made in this paper.
We can safely use the profile function, say $E \gtrsim 300$\,MeV,
where the difference between $pn$ and $pp$
cross sections are not significant.
The validity of adopting
the averaged $NN$ profile function here 
is discussed in Appendix A.
We distinctively use the $pn$ and $pp$ profile 
functions~\cite{Ibrahim08} when
more realistic cases are considered in Sec.~\ref{diffpn.sec}.
We do not include the spin-orbit term
  in the profile function~\cite{Ibrahim08}.
  As long as focusing on the analysis of
  the elastic scattering differential cross sections at the forward angles,
  this effect is small at the peak positions~\cite{Kohama04},
  whereas the cross sections at
  the diffraction minima are significantly influenced
  by the spin-orbit interaction~\cite{Alkhazov78}.
  In fact, as shown in Ref.~\cite{Horiuchi16},
  the elastic scattering differential cross sections data
  at the forward angles are fairly well reproduced 
  using the profile function (\ref{profile.eq}) without the spin-orbit term.
The elastic and inelastic Coulomb contributions are ignored
since the effects are negligible
in the proton-nucleus scattering~\cite{Horiuchi16}.

\section{Results}
\label{results.sec}

In this section, we show the results of the analyses 
explained in the previous section.

\subsection{Elastic scattering differential cross section 
and nuclear surface distribution}
\label{Fermi.sec}

\begin{figure}[ht]
\begin{center}
    \epsfig{file=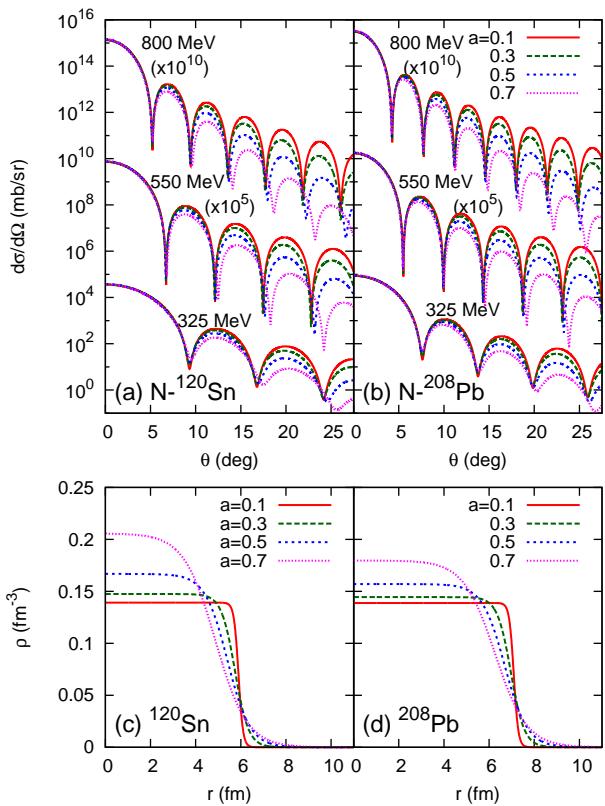, scale=1}
    \caption{Elastic scattering differential cross sections
      of (a) $N$-$^{120}$Sn and (b) $N$-$^{208}$Pb systems
      calculated with 2pF density distributions
      at 325, 550, and 800\,MeV. The cross sections are multiplied
      by $10^5$ and $10^{10}$ for those at 550 and 800\,MeV, respectively.
      The lower panels plot the corresponding
      2pF density distributions of
      (c) $^{120}$Sn and (d) $^{208}$Pb, respectively,
      with various diffuseness parameter $a$. All density distributions 
      give the same root-mean-square radius.}
 \label{Fermi.fig}
\end{center}
\end{figure}

Here we discuss how much information the elastic scattering
cross sections of the first
diffraction peak has on the nuclear surface.
In the present work,
we are interested in medium to heavy nuclei
whose central densities are well saturated.
It would be reasonable to assume
a two-parameter Fermi (2pF) function
as an approximate nuclear density distribution:
\begin{align}
  \rho(r)=\frac{\rho_0}{1+\exp\left(\frac{r-R}{a}\right)},
\label{Fermi.eq}
\end{align}
where $\rho_0$, $R$, and $a$ are the central density,
radius, and diffuseness parameters, respectively.
For given $R$ and $a$, $\rho_0$ is uniquely determined
by the normalization condition: $4\pi \int_0^\infty \rho(r) r^2 dr=A$,
where $A$ is the mass number of a nucleus.
The root-mean-square (rms) matter radius
can be calculated by
\begin{align}
r_m=\sqrt{\left<r^2\right>}= \sqrt{\frac{4\pi}{A}\int_0^\infty dr\,r^4\rho(r)}.
\end{align}
We note that the limit $a\to 0$ in Eq.~(\ref{Fermi.eq})
results in a sharp-cut square-well density distribution
with a radius $R=\sqrt\frac{5}{3}r_m$.

We perform the Glauber model calculation with
the 2pF density distribution of Eq.~(\ref{Fermi.eq}).
 The rms radius of the 2pF density distribution
  is set to follow the empirical
  rms radius $r_m=\sqrt{\frac{3}{5}}1.2A^{1/3}$~\cite{BM}
  as a convenient choice
  so that the radius parameter $R$ is determined for each given $a$.
Note that the resultant $R$ is in general
different from the radius
obtained by the sharp-cut square-well density distribution,
$\sqrt\frac{5}{3}r_m$.

\subsubsection{Nuclear radius $\leftrightarrow$
  First peak position of the elastic scattering 
  differential cross section}

First, we discuss the relation between the nuclear radius
and the scattering angle at the first peak position
of the elastic scattering differential cross section.
Figure~\ref{Fermi.fig} plots the elastic scattering differential
cross sections of $N$-$^{120}$Sn and $N$-$^{208}$Pb systems incident at
325, 550, and 800\,MeV with various diffuseness parameter, $a$.
Corresponding 2pF density distributions
are also plotted in the lower panels of Fig.~\ref{Fermi.fig}.
Focusing on the first diffraction peak,
all the cross sections are peaked
at almost the same scattering angle for all incident energies 
under consideration.
Since the BS model works well in the nucleon-nucleus scattering,
the one-to-one correspondence between the peak position
and the nuclear radius can be naturally understood,
remarking that all the 2pF density distributions
give the same rms radius.

\begin{figure}[t]     
  \begin{center}
\epsfig{file=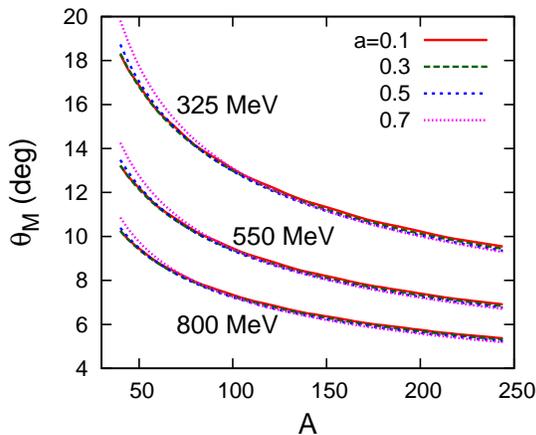, scale=1.2}    
\caption{Scattering angles at the first peak positions
  of the elastic scattering differential 
  cross sections incident
  at 325, 550, and 800\,MeV as a function of mass number.}
    \label{angle1stpeak.fig}
  \end{center}    
\end{figure}

For more quantitative discussions,
we display, in Fig.~\ref{angle1stpeak.fig},
the scattering angles at the first peak position, $\theta_M$,
as a function of mass number.
Incident energies of 325, 550, and 800\,MeV are chosen.
We again confirm that 
the first peak positions do not depend on the diffuseness parameter
of the 2pF density distribution.
The peak position is determined 
mostly by the nuclear radius.
For a small $A\lesssim 70$, the $\theta_M$ values show
some dependence on $a$, especially with large $a=0.7$\,fm.
Since we assume the 2pF distribution,
in the case of small $A$, i.e., small $R$,
large diffuseness parameter substantially affects
the density profile at small distances, which is,
a large increase of the central density
as already seen in the lower panels of Fig.~\ref{Fermi.fig}.

\subsubsection{Nuclear diffuseness $\leftrightarrow$
  Magnitude of elastic scattering differential cross section
  at the first peak position}

Second, we discuss the relation between the nuclear diffuseness
and the magnitude of the elastic scattering
differential cross section at the first peak position.
In Fig.~\ref{Fermi.fig}, it is interesting to note that
the elastic scattering differential cross section values
at the first peak position are mostly determined by $a$.
The authors of Ref.~\cite{Amado80} pointed out
a relation between the nuclear surface
diffuseness and the elastic scattering differential cross section,
in which the cross section at the first peak position
is enhanced with smaller nuclear diffuseness.
The calculated cross sections actually show
a larger value at the first peak position with smaller $a$.

\begin{figure}[t]     
  \begin{center}
\epsfig{file=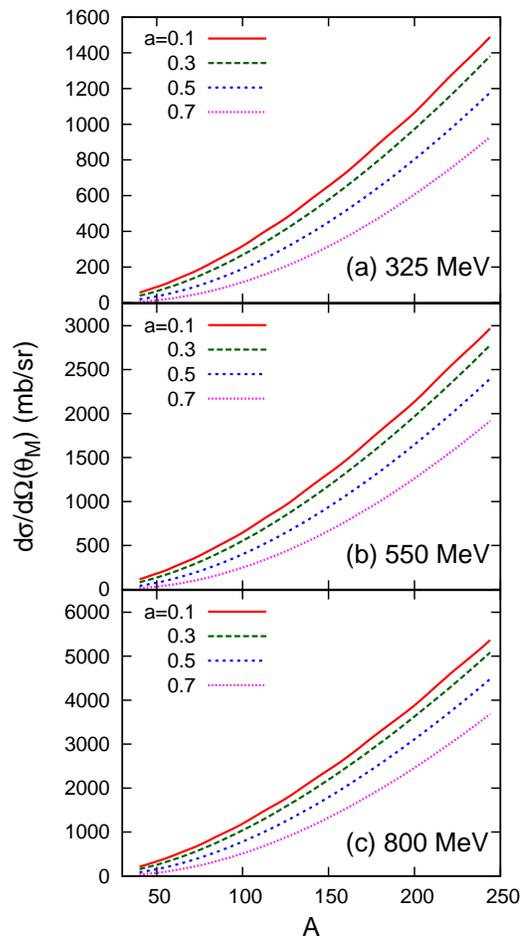, scale=1.1}    
    \caption{Elastic scattering differential cross sections
      at the first peak position incident at
      (a) 325, (b) 550, and (c) 800\,MeV
      as a function of mass number.}
    \label{cs1stpeak.fig}
  \end{center}    
\end{figure}

Figure~\ref{cs1stpeak.fig} plots
the magnitude of the elastic scattering
differential cross section at the scattering
angle of the first peak position $\theta_M$
as a function of mass number.
Incident energies of 325, 550, and 800\,MeV are chosen.
The cross section significantly decreases with increasing $a$
which would easily be distinguished by measurement.
Though the sensitivity to $a$ becomes a little
bit less at 800\,MeV,
higher incident energy gives larger cross sections.
We find that for a given $A$, i.e., rms radius,
the cross sections with different $a$ do not intersect
each other for all the incident energies,
indicating that the $R$ and $a$ parameters
of the 2pF density distribution can uniquely be determined
if the elastic scattering differential cross section is measured at
the first peak position.

\subsection{Scattering amplitude at the first and second peak positions}
\label{samp.sec}

We have seen so far
that the nuclear ``diffuseness'' information can be extracted
from the elastic scattering differential cross section 
at the first peak position.
In this subsection, we discuss what the incident nucleon actually probes.
To answer this, we calculate the scattering amplitude
of the differential cross section
at the first peak position $\theta_M$
as a function of the impact parameter $b=|\bm{b}|$
\begin{align}
  f_1(\bm{b})=\frac{iK}{2\pi}e^{-i\bm{q}_M\cdot\bm{b}}
  \left(1-e^{i\chi(\bm{b})}\right),
\label{scatampb.eq}
\end{align}
where $\bm{q}_M\cdot\bm{b}=2 K b \sin\frac{\theta_M}{2}$,
and the relation to the scattering amplitude
at the first peak position 
\begin{align}
F(\theta_M)=\int f_{1}(\bm{b})\,d\bm{b}.
\end{align}

\begin{figure}[ht]
  \begin{center}
    \epsfig{file=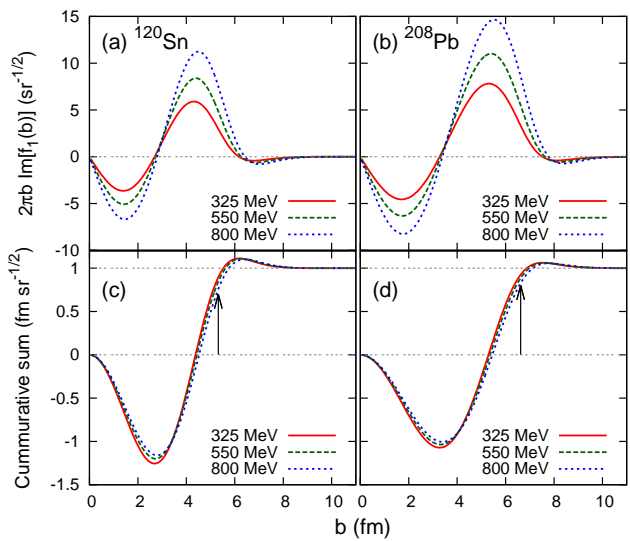, scale=0.9}    
    \caption{(Upper) Imaginary part of the spatial distribution
      of the scattering amplitude, and (Lower) its cumulative sum
      at the first peak of the elastic scattering differential cross section
      of (a) (c) $^{120}$Sn and (b) (d) $^{208}$Pb,
      as a function of impact parameter $b$,
      incident at 325, 550 and 800\,MeV. An arrow indicates
      the half density radius of (c) $^{120}$Sn and (d) $^{208}$Pb, respectively.
      See text for details.
    }
    \label{amplitude.fig}
  \end{center}
\end{figure}

Figure~\ref{amplitude.fig} plots
the imaginary part of the spatial distribution
of the scattering amplitude of Eq.~(\ref{scatampb.eq})
for $^{120}$Sn and $^{208}$Pb 
and its cumulative sum
defined by $2\pi b\int_0^b {\rm Im} f_{1}(b^\prime) db^\prime
/{\rm Im}F(\theta_M)$,
at various incident energies as a function of impact parameter $b$.
The real part is not shown because it is small.
The diffuseness parameters are set commonly to 
an empirical value $a=$0.54\,fm~\cite{BM}.
The half density radius $\rho(R_{h})=\rho_0/2$ for each nuclide
is indicated by an arrow.
All curves exhibit positive and negative peaks
inside the nuclear half radius.
The cumulative sum of $^{120}$Sn ($^{208}$Pb)
indicates that the amplitude up to $\sim 4.5$ \,fm ($\sim 5.5$\,fm)
does not contribute to the integrated scattering amplitude
as they are canceled out through the integration over $b$.
As a result, only the scattering amplitude
at around $R_h$ is contributed to the cross sections
in such a special kinematic condition.
The nucleon-nucleus cross section at
the first peak position can be
a useful observable to extract the density profiles
at around the half density radius.

\begin{figure}[ht]
  \begin{center}
    \epsfig{file=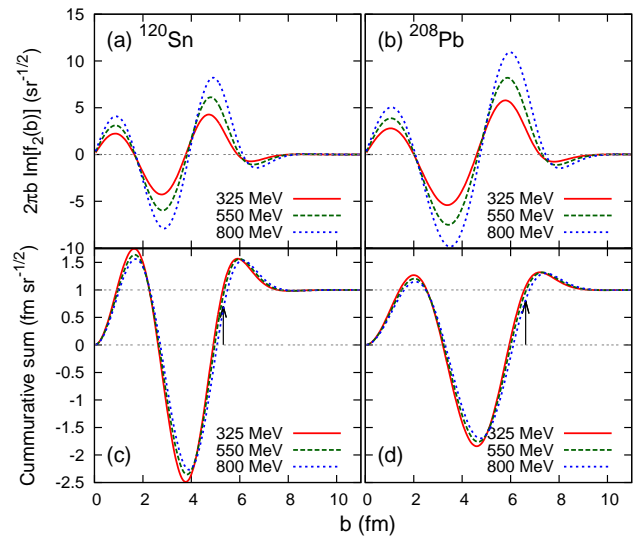, scale=0.9}    
    \caption{Same as Fig.~\ref{amplitude.fig} but at the second peak
      of the elastic scattering differential cross section.
    }
    \label{amplitude2nd.fig}
  \end{center}
\end{figure}

We comment on what information can be obtained
in the higher-order diffraction peak.
We see, in Fig.~\ref{Fermi.fig}, that peak positions of
the higher-order diffraction
are shifted to larger scattering angles with increasing $a$,
implying different sensitivity to the nuclear density profile.
Figure~\ref{amplitude2nd.fig} plots the same quantity
as Eq.~(\ref{scatampb.eq}) but at the second peak position,
$f_{2}(\bm{b})$. The spatial distribution
allows one more node and varies more rapidly with increasing $b$.
The cumulative sum also oscillates and shows some contribution,
reaching at the tail region which is a bit distant
from the half density radius.
The cross section at the second peak
would have some other information on
the density profile than that at the first peak.

Further investigation would be interesting
since it is useful to extract 
a higher order of the density profile beyond the half density radius
which characterizes weakly bound systems, e.g., halo nuclei
but it is beyond the scope of this paper.

\subsection{Extraction of nuclear radius and ``diffuseness''}
\label{extdiff.sec}

Thus far, we have discussed that
the nuclear radius and diffuseness information is embedded
in the first peak position and its magnitude
of the elastic scattering cross section.
In this subsection, we demonstrate
how we can extract the nuclear ``diffuseness'' as nuclear structure
information when the nucleon-nucleus elastic scattering differential cross
sections are given. For this purpose, we employ
general proton and neutron density distributions obtained
by a microscopic mean-field model
as inputs to the Glauber model.

We take the density distributions of Ca, Ni, Zr, Sn, Yb, and Pb isotopes
obtained by the Skyrme-Hartree-Fock (HF) + BCS method~\cite{Ebata10}
used in Refs.~\cite{Horiuchi16,Horiuchi17b}
(One can also take them from the theoretical database~\cite{Ebata18}).
The calculation was performed self-consistently
in a three-dimensional Cartesian mesh, in which
any nuclear deformation can be taken into account.
The density distribution in the laboratory frame is obtained
by taking an average on the angles as in Ref.~\cite{Horiuchi12}.
We remark that the theoretical justification of this treatment
was made in Ref.~\cite{Sumi12}.
The Skyrme-type effective interaction (SkM*~\cite{SkMs})
with a monopole-type pairing interaction
is employed as detailed in Refs.~\cite{Ebata14,Ebata17}.
The SkM* parameter set is superior to describe the nuclear deformation.
For example, kink behavior due to the nuclear deformation in the
total reaction cross sections of neutron-rich Ne and Mg isotopes
are reproduced very well with the help
of the Glauber model~\cite{Horiuchi12,Horiuchi15}.
  
To deduce the ``diffuseness'' of the realistic density distribution
through reaction data,
we calculate the elastic scattering differential cross sections
with the 2pF density distribution of Eq.~(\ref{Fermi.eq}).
Regarding that those calculated cross sections
with the HF+BCS density distributions 
are ``experimental data'', we determine
the $R$ and $a$ in the 2pF density distribution
in such a way that the calculated elastic scattering 
differential cross section
matches the first peak position
as well as its magnitude of
the elastic scattering differential cross section obtained
by the ``experiment''.
To assure the accuracy of the extracted
$R$ and $a$, we confirm that our cross section
calculations are converged within four digits.

\subsubsection{Uncertainties of the extraction}

\begin{figure*}[ht]
  \begin{center}
    \epsfig{file=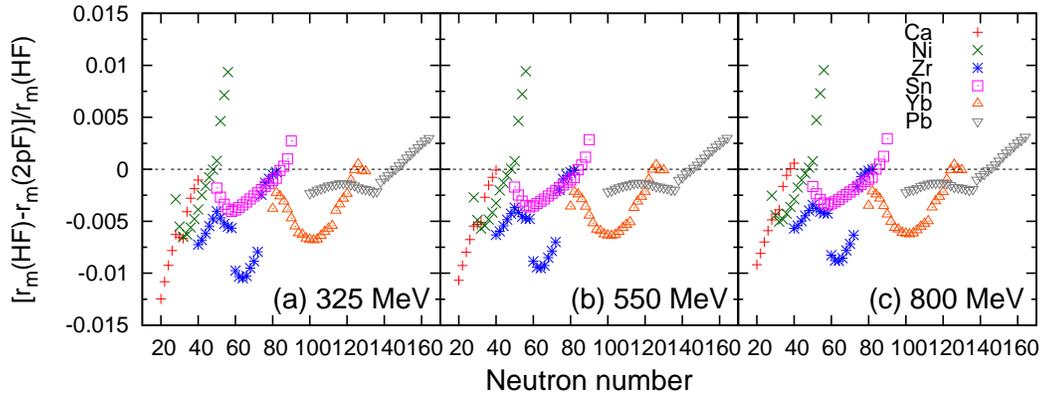, scale=1.2}    
    \caption{Relative deviations of the rms radius
      obtained by the HF+BCS method and the deduced rms radii
      incident at (a) 325, (b) 550, and (c) 800\,MeV.
      }
    \label{corr.fig}
  \end{center}    
\end{figure*}

Following the procedure mentioned above,
we determine the parameters in the 2pF density distributions
for each isotope and each incident energy.
In this subsection, we evaluate the robustness of this analysis.

Figure~\ref{corr.fig} plots the relative deviations
of the rms radius obtained by the HF+BCS density, $r_m$(HF),
and that extracted from this analysis,
$r_m$(2pF) for Ca, Ni, Zr, Sn, Yb and Pb isotopes.
These results indicate the difference between realistic HF+BCS
  and simple 2pF density distributions which will be commented
  later in this section and Sec.~\ref{diffuseness.sec}.
  Here we choose three incident energies 
  (a) 325, (b) 550, and (c) 800\,MeV.
The deviations are typically less than 1\% for all the incident energies.
The extracted rms radius agrees very well with the ``correct'' rms radius
and is successfully obtained from this analysis.
The 2pF density distribution
  can be a reasonable approximation to simulate the
  realistic density distributions of medium- to heavy-mass nuclei.

\begin{figure}[ht]
  \begin{center}
    \epsfig{file=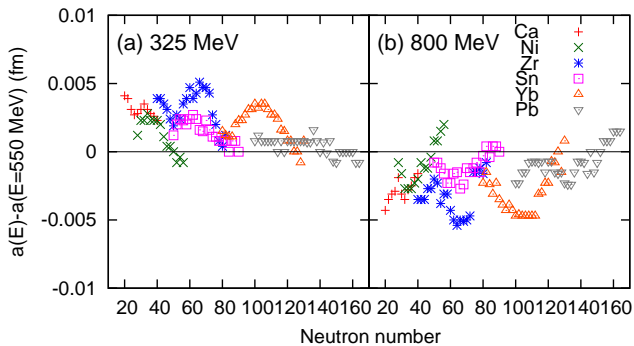, scale=1.05}    
    \caption{Deviations of the nuclear ``diffuseness''
      deduced at (a) 325 and (b) 800\,MeV from that at 550\,MeV.
      See text for details.}
    \label{diffHF-enedep.fig}
  \end{center}
\end{figure}

For the extraction of the diffuseness parameter $a$,
we display, in Fig.~\ref{diffHF-enedep.fig},
the deviations of the diffuseness parameter $a$
at 325 and 800\,MeV from that at 550\,MeV.
The incident-energy dependence is small at most by $\sim0.005$\,fm.
Though some systematic errors exist which
  come from the difference between the realistic and 2pF density distributions,
  we can however determine, within this model approach,
  such an $a$ value in the accuracy of two digits.
The $a$ value, which is extracted in this way,
can be used as a measure of the surface diffuseness
for the realistic density distribution.

\begin{figure*}[ht]
  \begin{center}
    \epsfig{file=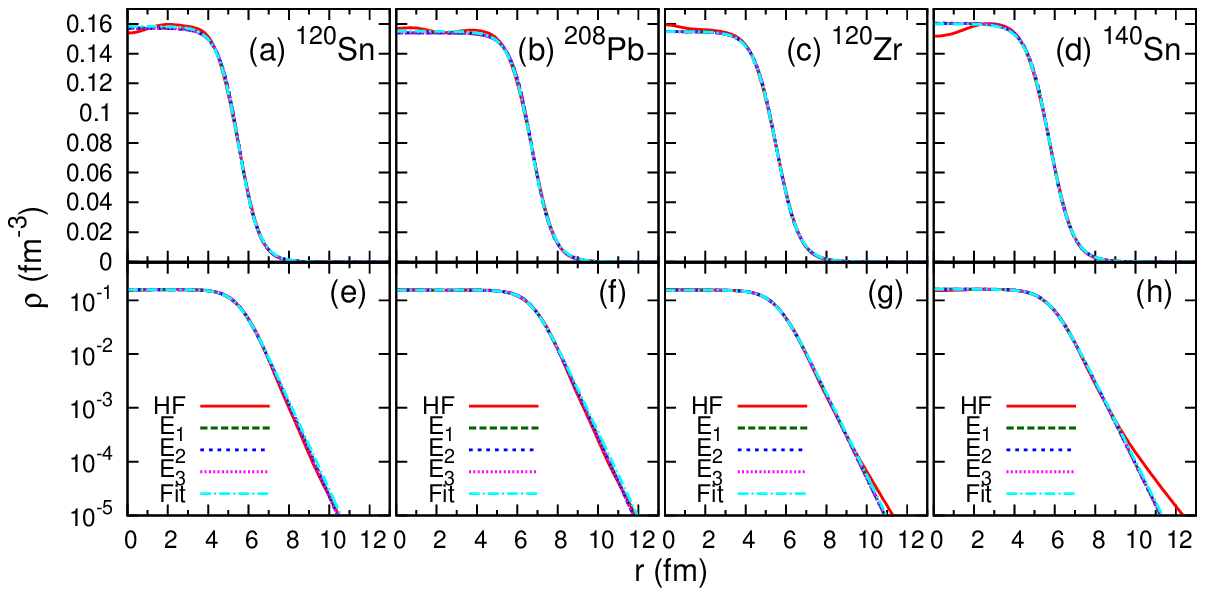, scale=1.4}    
    \caption{Comparison of
      HF+BCS density and deduced 2pF density distributions of
      (a) (e) $^{120}$Sn, (b) (f) $^{208}$Pb, (c) (g) $^{120}$Zr, and (d) (h)
        $^{140}$Sn at different incident energies, 
      $E_1=325$, $E_2=550$, and $E_3=800$\,MeV,
      drawn in (upper) linear and (lower) logarithmic scales.      
        'Fit' denotes the 2pF density distribution deduced directly from
    the HF+BCS density. See text for details.}
    \label{density.fig}
  \end{center}
\end{figure*}

Figure~\ref{density.fig} compares the HF+BCS density with
the 2pF distributions deduced from the first peak position
and its magnitude of the elastic scattering differential cross section.
The 2pF density distributions well simulate
HF one at around the nuclear surface at the three incident energies,
$E_1=325$, $E_2=550$, and $E_3=800$\,MeV.
However, we see some deviations beyond $\sim 9$\,fm
  for the neutron-rich Zr and Sn nuclei which cannot be expressed
  by a simple 2pF distribution.
  This trend can also be seen in the behavior of the relative
  deviations of the rms radius plotted in Fig.~\ref{corr.fig}.
  We also determine the 2pF density distribution
  by minimizing the root-mean-square deviation between the 2pF and
  HF+BCS density distributions.
  As plotted in Fig.~\ref{density.fig}, the resultant 2pF density distributions
  are almost identical with those extracted from the first peak of the
elastic scattering differential cross sections.
The extracted $R$ and $a$ values can be robust structure information
independently from the choice of the incident energy.

\subsubsection{Systematic trend of the nuclear ``diffuseness''}
\label{diffuseness.sec}

\begin{figure}[ht]
  \begin{center}
    \epsfig{file=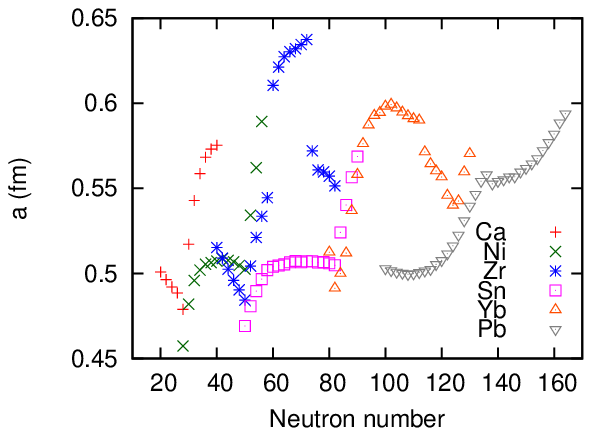, scale=1.2}    
    \caption{Nuclear ``diffuseness''
      deduced from the HF+BCS density distributions incident at 550\,MeV.}
    \label{diffHF.fig}
  \end{center}
\end{figure}

It is interesting to see the behavior of the nuclear
``diffuseness'' deduced
from the HF+BCS density distributions.
Figure~\ref{diffHF.fig} displays the deduced $a$ values at 550\,MeV.
The $a$ values are scattered around $a\sim 0.5$\,fm,
which is quite reasonable by remarking
the empirical value $\sim$0.54\,fm for stable nuclei~\cite{BM}.
As discussed in Ref.~\cite{Horiuchi17b},
the nuclear diffuseness is closely related to the width
of the nuclear surface that determines the surface tension
of the nuclear droplet. 
This clearly indicates some exotic nuclear structure,
such as nuclear deformation, and weakly bound orbits.
We note that the diffuseness parameters extracted in this paper
are that for the matter density distributions.
The neutron number dependence is somewhat
weaker than that of the neutron surface widths obtained
in Ref.~\cite{Horiuchi17b} because
the proton surface widths are small
and almost stay at a constant in
the neutron-rich isotopes.
Though it exhibits the weaker dependence
on the neutron number than that of the neutron diffuseness,
we can still see the structural information
on the exotic neutron-rich isotopes.
As expected, the $a$ values show local minima at the magic numbers.
The $a$ values exhibit sudden rises at
$N=50$ for Ca and Ni isotopes, and at $N=82$ for Sn isotopes,
in which weakly bound neutron orbits play 
a role~\cite{Ebata14, Horiuchi17a}.
Large $a$ values in the open shell regions
of Zr and Yb isotopes are due to the nuclear deformation,
similarly to the cases of
the Ne and Mg isotopes~\cite{Minomo11,Minomo12,Sumi12,Horiuchi12,Watanabe14}.
A systematic measurement of the elastic scattering 
differential cross sections
covered up to the first diffraction peak
will have of particular importance in order to reveal the evolution
of the exotic structure of unstable nuclei.

It should be noted that this method 
may not be applicable to very weakly bound systems,
such as halo nuclei, because
the density profile deviates considerably from
the 2pF assumption.
To get higher resolution of the density profile,
one may consider an analysis including higher-order
diffraction peaks with more general density distribution
other than the 2pF distribution.

\subsection{Black sphere estimate of the first peak position}
\label{BS.sec}

We have discussed that one can obtain
the rms radius and nuclear diffuseness simultaneously
from the first peak position and its magnitude of the
elastic scattering differential cross section.
It would be helpful to know the peak position before measurement
of the elastic scattering differential cross sections.
For this purpose, we investigate quantitative relation between
the first peak position and the total reaction cross section using
a concept of the strong absorption, i.e., 
the framework of the BS model~\cite{Kohama04}.

If a nucleus is a completely absorptive object
within a sharp-cut nuclear radius $a_{\rm BS}$,
the total reaction cross section reads exactly as
\begin{align}
\sigma_{\rm BS}=\pi a_{\rm BS}^2.
\label{BSrcs.eq}
\end{align}
Note that the same thing holds 
for the total elastic cross section as well.
The BS radius $a_{\rm BS}$ is obtained 
by the angle $\theta_M$ corresponding to the first diffraction
peak as~\cite{Kohama04}
\begin{align}
  a_{\rm BS}=\frac{5.1356\cdots}{2p\sin(\theta_M/2)},
\label{BSang.eq}
\end{align}
where $p$ $(=K)$ 
is the momentum between the two colliding particles.

As in Eq.~(\ref{BSrcs.eq}), the total reaction cross section is
directly related to the scattering angle of the first peak position, 
but, in reality, the total reaction cross section deviates
from the one obtained from Eq. (\ref{BSrcs.eq})
due to the nuclear transparency which comes from the surface diffuseness.
In the Glauber model, the total reaction cross section is calculated by 
\begin{align}
\sigma_R=\pi a_R^2=\int \left(1-|e^{i\chi(\bm{b})}|^2\right)\,d\bm{b},
\label{reacrad.eq}
\end{align}
where we see similarity to Eq.~(\ref{BSrcs.eq}) by introducing
a reaction radius $a_R$~\cite{Horiuchi14,Horiuchi16}.
In contrast, the $a_{\rm BS}$ is determined from the first peak 
position of
the calculated elastic scattering differential 
cross section~\cite{Kohama04}, 
using Eq. (\ref{BSang.eq}),
and obtain $\sigma_{\rm BS}$ by the formula (\ref{BSrcs.eq}).

\begin{figure*}[ht]
  \begin{center}
    \epsfig{file=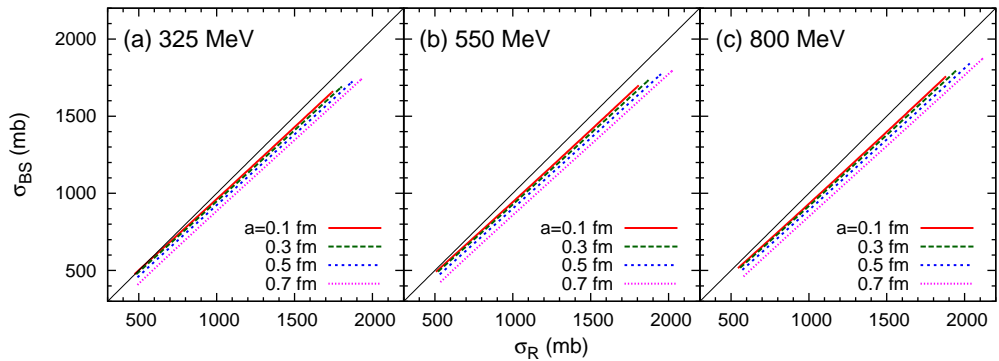, scale=1}
    \caption{Comparison of the total reaction cross sections
      obtained by the Glauber (horizontal axis) 
      and BS models (vertical axis).
    A solid-thin line indicates a $y=x$ line plotted to guide the eyes.}
    \label{bsvsr.fig}
  \end{center}
\end{figure*}

Figure~\ref{bsvsr.fig} compares
the total reaction cross sections incident at 325, 550, and 800\,MeV
obtained by the Glauber calculation 
and the BS estimate using Eq.~(\ref{BSrcs.eq}).
The 2pF density distributions are employed.
Since the BS model assumes the sharp-cut-square-well
nuclear surface and
complete absorption in $r\leq a_{\rm BS}$,
the deviation must include the information on the nuclear diffuseness
or nuclear transparency at around the surface.
If the nucleon-nucleus scattering is the ideal black sphere, 
all results will be on a $y=x$ line drawn in this figure.
However, some deviation is found
indicating the difference between the BS model
and actual proton-nucleus scattering.
Since more nucleons at around the nuclear surface
contribute to the scattering process,
the deviation becomes larger and larger
with increasing nuclear size and diffuseness,
which is typically by $\sim 5$\%, at most by
$\sim$10\% at Pb isotopes in the ranges of the standard diffuseness
parameters $a=0.5$-0.6\,fm.

We note that the energy dependence of the deviation
cannot be explained simply by the magnitude
of the total nucleon-nucleon cross section
but it comes from the difference of the nucleon-nucleon scattering processes
or the profile functions in the present model, which
is reflected in the nuclear transparency at around the surface.
In fact, we confirm that, the BS estimate gives the same cross section
at any incident energy when the same profile function is used.

\begin{figure*}[ht]
  \begin{center}
    \epsfig{file=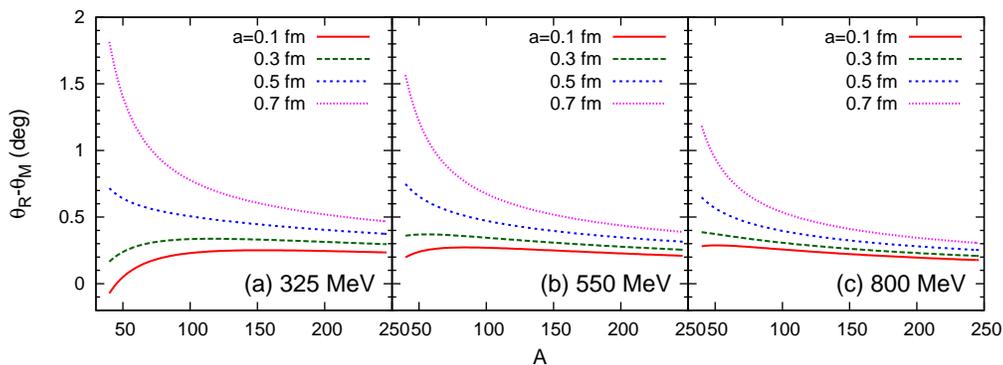, scale=1}
    \caption{Difference of the scattering angles of the first 
    peak position
    obtained by the Glauber calculation and the BS estimate
    incident at (a) 325, (b) 550, and (c) 800\,MeV.}
    \label{tbsvsr.fig}
  \end{center}
\end{figure*}

We have shown that
the BS model explains more than 90\% of the nucleon-nucleus scattering.
Moreover, it is practically important to know how accurate we can obtain
the first peak position if one converts 
the total reaction cross section
to the scattering angle of the first peak position
using the relation of Eq.~(\ref{BSang.eq}).
Figure~\ref{tbsvsr.fig} displays
the difference between $\theta_M$ and $\theta_R$.
The former can be obtained directly
from the first peak position of
the elastic scattering differential cross sections,
and the latter can be calculated 
by converting the relation
\begin{align}
a_R= \sqrt{\frac{\sigma_{R}}{\pi}}=\frac{5.1356\cdots}{2p\sin(\theta_R/2)},
\label{BSest.eq}
\end{align}
through Eq.~(\ref{reacrad.eq}). Again, we note that
if the nucleon-nucleus scattering is completely absorptive,
the difference must be zero.
Despite the fact that the $\sigma_{\rm BS}$ slightly
deviates from $\sigma_R$ as shown in Fig.~\ref{bsvsr.fig},
the differences of those scattering angles
appear to be small, only within a few degrees.
Therefore, Eq.~(\ref{BSest.eq}) works well
for the estimation of the first peak of the nucleon-nucleus diffraction,
and thus the total reaction cross section can be complementary
information to set up scattering angles to be covered by measurement.

\subsection{Diffuseness of proton and neutron surfaces}
\label{diffpn.sec}

To extract detailed structure information of unstable nuclei,
separation of proton and neutron diffuseness is important
because the neutron diffuseness is expected to be more sensitive
to the ground state structure of neutron-rich isotopes
as it is dominated by the neutron motion at the nuclear surface.
As was done in Refs.~\cite{Terashima08,Zenihiro10},
neutron distributions of stable nuclei can be extracted
from the proton-nucleus elastic scattering measurements
using a known proton density distribution but it is
in general unknown for unstable nuclei.
Here, we discuss the possibility 
of making use of the incident energy dependence
of the $pn$ and $pp$ total cross sections
as utilized in Refs.~\cite{Horiuchi14,Horiuchi16}.
We extend that idea in order to extract both the proton and neutron
surface diffuseness and radii.

We respectively assume the 2pF density distributions (\ref{Fermi.eq})
for proton and neutron and determine these four parameters
in such a way as to reproduce the first peak positions 
and their differential elastic scattering 
cross sections at low and high incident energies.
A realistic profile function~\cite{Ibrahim08}, which
differs for $pp$ and $pn$, is used for the Glauber calculation.
Table~\ref{diff.tab} lists the extracted
diffuseness parameters and rms radii for proton and neutron.
Stable $^{120}$Sn, $^{208}$Pb and neutron-rich $^{132}$Sn 
isotopes are chosen as the examples.
We choose several sets of two incident energies among 200, 300, 550,
and 800\,MeV.
For $^{120}$Sn and $^{208}$Pb, extracted diffuseness parameters 
are scattered although the rms radii are converged within $\sim 0.5\%$. 
In such cases where the proton and neutron surfaces are located
at almost the same position,
the separation of the proton and neutron surface profiles might be difficult,
whereas, in case of $^{132}$Sn,
all extracted values are consistent with each other. This method can be used
to extract the information on the proton and neutron surfaces
from the proton-nucleus elastic scattering
in the inverse kinematics,
although the application of the method is limited
only to such neutron(proton)-rich systems
that the surfaces of the proton and neutron density distributions
are well separated $\gtrsim 0.2$\,fm.

 It should be noted
  the separation of the proton and neutron density distributions
  will be better by employing the cross sections
  at lower incident energies
  $\lesssim 200$\,MeV, where the $pn$ total cross section becomes
  much larger than that of the $pp$ one.
  As demonstrated in Ref.~\cite{Horiuchi16}, the adopted Glauber model
  is reliable for a wide range of the incident energies,
  even at few tens of MeV.
  However, with lowering the incident energy,
  in-medium effects such as
  Pauli blocking and Fermi-motion would be important and
  may modify the parameters of the free $NN$ profile
  function~\cite{Hussein91, Takechi09}.
  Implementing these effects will be interesting
  for further improvement of the adopted Glauber model.

\begin{table}[b]   
\begin{center}
  \caption{Diffuseness parameters and rms radii for
    matter ($r_m$), neutron ($r_n$) and proton ($r_p$) extracted
    from the HF+BCS density distributions of $^{120,132}$Sn and $^{208}$Pb.
    at several two choices of incident energies  
    among 200, 300, 550, and 800\,MeV ($E_L<E_H$). 
    Units are given in MeV and fm for energy and length, respectively.}
    \begin{tabular}{ccccccccccccc}
      \hline\hline
      Nuclide && $(E_L,E_H)$&&$r_m$&&$r_n$&& $a_n$ &&$r_p$ &&$a_p$\\
      \hline
      $^{120}$Sn
      &&(200,300)           &&4.691&&4.725&&0.455&&4.645&&0.619\\      
      &&(200,550)           &&4.686&&4.720&&0.506&&4.639&&0.507\\
      &&(200,800)           &&4.685&&4.724&&0.470&&4.629&&0.525\\
      &&(300,550)           &&4.683&&4.708&&0.543&&4.648&&0.455\\
      &&(300,800)           &&4.683&&4.713&&0.543&&4.640&&0.448\\
\cline{2-13}
      && HF+BCS             &&4.662&&4.723&&     &&4.576&&     \\
\hline
$^{208}$Pb
      &&(200,300)           &&5.580&&5.604&&0.492&&5.542&&0.604\\      
      &&(200,550)           &&5.575&&5.608&&0.532&&5.424&&0.507\\
      &&(200,800)           &&5.574&&5.613&&0.542&&5.514&&0.479\\
      &&(300,550)           &&5.571&&5.592&&0.558&&5.538&&0.463\\
      &&(300,800)           &&5.570&&5.603&&0.557&&5.519&&0.458\\
\cline{2-13}
&& HF+BCS             &&5.551&&5.617&&     &&5.448&&     \\
\hline
$^{132}$Sn
      &&(200,300)           &&4.821&&4.851&&0.539&&4.776&&0.448\\      
      &&(200,550)           &&4.823&&4.856&&0.539&&4.765&&0.445\\
      &&(200,800)           &&4.822&&4.875&&0.535&&4.723&&0.443\\
      &&(300,550)           &&4.818&&4.844&&0.539&&4.779&&0.446\\
      &&(300,800)           &&4.820&&4.852&&0.537&&4.763&&0.445\\
\cline{2-13}
&& HF+BCS             &&4.802&&4.890&&     &&4.656&&     \\
\hline\hline
  \end{tabular}
  \label{diff.tab}
\end{center}
\end{table}

\section{Conclusion}
\label{conclusion.sec}

In order to see how much we can extract information
on density profiles of unstable nuclei at around the nuclear surface,
we have performed a ``numerical experiment'' using
theoretically obtained elastic scattering differential cross sections
of high-energy nucleon-nucleus scattering incident at
a few to several hundreds of MeV.
The high-energy nucleon-nucleus collision is described
by the Glauber model starting from the nucleon-nucleon 
total cross sections.

We have demonstrated that the elastic scattering
differential cross section at the first diffraction peak reflects
the nuclear density profile at around the half density radius.
This can be understood naturally by extending
the idea of the black sphere (BS) model
offering the one-to-one correspondence between
the nuclear radius and the diffraction peak.
The deviation of the BS picture
from the actual nucleon-nucleus scattering
exhibits the role of the nuclear transparency
due to the diffused nuclear surface.
We have understood that the BS model is accurate but 
accompanied with typically $\sim$5\% uncertainties
in medium-mass nuclei,
at most $\sim$10\% uncertainties in Pb isotopes.

Towards the application to studies of unstable nuclei,
since the elastic scattering differential cross section
data at large scattering angles are
hardly obtained, we restrict ourselves to have only two observables,
the first peak position and
its magnitude of the elastic scattering differential cross section.
Assuming that the two-parameter Fermi (2pF) density distribution
as a fitting density,
we can uniquely determine the two parameters which result in
the root-mean-square (rms) radius and nuclear diffuseness.
A systematic ``numerical experiment''
  is performed using realistic density distributions
obtained by a microscopic mean-field model.
The accuracy of the extraction
does not depend much on the incident energy.
Though the simple 2pF form is assumed
  as an approximate density distribution, 
the rms radius can be determined within $\sim$1\%,
and the extracted nuclear ``diffuseness'' is robust
structure information
that reflects interesting surface profiles
on the exotic nuclei such nuclear deformation and shell evolution.

Since we only need the cross section at the first peak position,
this method has an great advantage to apply to measurements
in the inverse kinematics, in which the scattered particles
are concentrated at the forward angles.
The nuclear structure is actually reflected
  in the nuclear density profile at the surface, that is,
  the nuclear ``diffuseness''.
A systematic measurement along this direction is interesting
to understand structure changes of unstable nuclei
played by excess neutrons.

A prescription of separating the proton and neutron
radii and diffuseness is also given by making use
of the incident energy dependence 
for the case of a proton target.
If one measures 
the elastic scattering differential cross sections
at the first diffraction peak
at low ($E\lesssim 300$\,MeV) 
and high ($E\gtrsim 500$\,MeV) incident energies,
one can extract the surface diffuseness and the rms radii
of proton and neutron separately.
Though the method has limitation that can only be applied to
a nucleus with large neutron-skin thickness,
it will be useful to extract the structure information
of neutron- and proton-rich unstable nuclei.
  We note, however, some simplifications of the model
  are made in this analysis.
  In order to obtain more precise information on the density profile,
  we need a further study to quantify the systematic error of this analysis
  with a more elaborated model which includes many-fermion correlations
  as well as using a more flexible input density distribution.

We should point out here that the discussion extended 
in this paper must be applicable to the analyses of 
electron-nucleus elastic scattering differential cross sections, 
particularly for unstable nuclei, 
because the theoretical structure is quite similar to 
the proton scattering case~\cite{Amado80}, 
although the interaction is quite different.
By focusing on the first peak angle and its magnitude of 
the elastic scattering differential cross section, 
one may have a chance to determine the radius of the proton density 
distribution and its diffuseness for an unstable nucleus 
simultaneously. 
We believe that this is very important, but leave it for 
our future studies~\cite{scrit2017}.

\acknowledgments

The authors thank K. Iida for valuable communications
and S. Ebata for sending us numerical data
of density distributions obtained by the HF+BCS calculation.
This work was in part supported by JSPS KAKENHI Grant Numbers
  18K03635 and 18H04569.

\appendix

\section{Tests of averaged $NN$ profile functions}

\begin{figure}[t]   
  \begin{center}
    \epsfig{file=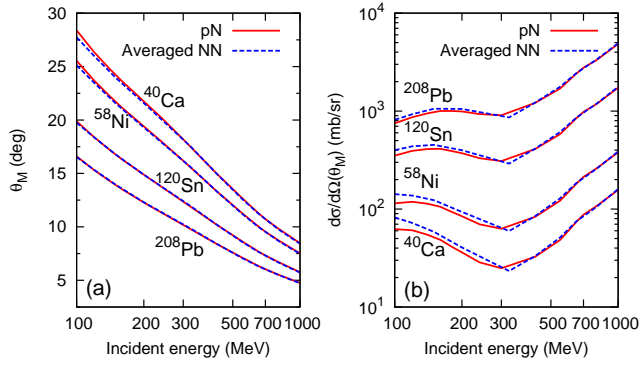, scale=0.9}
    \caption{(a) Scattering angles of the first peak position
    and (b) its magnitude of the elastic scattering differential 
    cross section at the peak position
      with the $pN$ and averaged $NN$ profiles functions.
      The HF+BCS density distributions 
      with the SkM* interaction~\cite{Horiuchi16,Horiuchi17b} are used.}
    \label{Angle-test.fig}
  \end{center}    
\end{figure}

In this appendix, we test the validity of 
the average procedure of $pn$ and $pp$ total cross section
in the profile functions~\cite{Horiuchi07}
for nucleon-nucleus systems employed in the analysis of this paper.
We calculate the first peak positions, $\theta_M$
  and its magnitude of 
  the elastic scattering differential cross section.
The proton and neutron density distributions
obtained by the HF+BCS method using
the SkM* effective interaction~\cite{Horiuchi16,Horiuchi17b}
are employed. 
Figure~\ref{Angle-test.fig} compares
  the results calculated with the $pN$~\cite{Ibrahim08}
  and averaged $NN$~\cite{Horiuchi07} profile functions
  at various incident energies.
We find that the peak positions do not depend on the choice of
the profile functions, while some differences are found
in the cross sections at $\theta_M$ at incident energies
lower than $\sim$300\,MeV, where the difference of
the $pn$ and $pp$ total cross section becomes significant.
We can safely use the averaged $NN$ profile function
for the scattering at $\gtrsim$300\,MeV,
the $pN$ profile functions should be used for quantitative
discussions of the proton-nucleus scattering at the lower energies.

\end{document}